# REFOCUS: CURRENT & FUTURE SEARCH INTERFACE REQUIREMENTS FOR GERMAN-SPEAKING USERS

Maximilian Speicher[1], Andreas Both and Martin Gaedke[1]
[1]*Technische Universität Chemnitz, 09111 Chemnitz*

**ABSTRACT**

While smartphones are widely used for web browsing, also other novel devices like Smart TVs become increasingly popular. Yet, current interfaces do not cater for the newly available devices beyond touch and small screens, if at all for the latter. Particularly search engines—today's entry points of the WWW—must ensure their interfaces are easy to use on any web-enabled device. We report on a survey that investigated (1) users' perception and usage of current search interfaces, and (2) their expectations towards current and future search interfaces. Users are mostly satisfied with desktop and mobile search, but seem to be skeptical towards web search with novel devices and input modalities. Hence, we derive *REFOCUS*—a novel set of requirements for current and future search interfaces, which shall address the demand for improvement of novel web search and has been validated by 12 dedicated experts.

## 1. INTRODUCTION

By the end of 2014, the number of mobile users exceeded the number of those using desktop PCs.[1] Moreover, the sales numbers of web-enabled novel devices total over 300 million units.[2] Hence, there is a strong shift towards devices beyond desktop PCs that provide options to browse the web, which include smart phones, tablet PCs, video game consoles and smart glasses, to name only a few. These *novel devices* introduce a new variety of input (touch, gestures, speech etc.) and output modalities (dominated by images, video, speech, 3D etc.) that go beyond what is possible with traditional keyboard, mouse and mid-sized computer screens[3]. This trend calls for novel, cutting-edge approaches towards web interfaces to adapt to the new circumstances and opportunities. However, from own observations we know that in today's industry, focus is still often on desktop PCs alone.

Search engines have become *the* main entry point to the WWW for many users. Particularly, three search engines are ranked among the top five *Alexa top sites*[4]. Also, the default start pages of web browsers on all modern devices feature prominent search boxes. This has led to users more and more “us[ing] web search engines as a replacement for web page bookmarking” (Weber and Jaimes, 2011). Also, Weber and Jaimes (2011) found that “the most frequent queries are navigational where the user might just as well type the query in the browser's address bar to re-find a URL from his browsing history”. Hence, for the user, URLs are only a mere intermediate step, which makes it highly necessary to provide usable search interfaces for the whole range of web-enabled devices that are available nowadays.

In this paper, we present—to the best of our knowledge—the first empirical study that results in a *requirements specification for both, current and future search interfaces*. The study was intended as *exploratory* research, which means that we do not try to answer a specific research question, but rather want to “get a feel for” users’ perceptions of and expectations towards current and future search engines. Our ten key findings show that users are generally satisfied with search experience on desktop PCs and mobile devices. Yet, they are rather skeptical towards novel devices and input modalities beyond touch. Users prefer input via regular keyboards and touch while tendentially disapproving speech recognition and motion-sensing techniques for web search. From our findings, we derived eleven requirements for any existing or future search interface to support developers and designers. These requirements have been reviewed and verified by 12 dedicated experts. Since the underlying survey was conducted almost exclusively with German-speaking users, the key findings and requirements are only valid for that specific group, as cultural differences have to be considered.

[1] http://www.businessinsider.com/mobile-will-eclipse-desktop-by-2014-2012-6 (July 20, 2016).
[2] http://goo.gl/bKshhP (July 20, 2016).
[3] In the context of this paper, we define the term NOVEL DEVICE as follows: *All types of devices that go beyond desktop PCs, laptops and commonly used smart phones and tablet PCs. The latter refers to the tier of devices that is currently represented by* iPhone, iPad *and alike. Examples for novel devices are smart TVs, cutting-edge video game consoles and tabletops.*
[4] http://www.alexa.com/topsites (July 20, 2016).

In the following, we describe a representative selection of related research (Sec. 2) and introduce the underlying survey (Sec. 3), the findings of which are described in Sec. 4. Sec. 5 defines and validates REFOCUS, our new requirements specification, before giving concluding remarks in Sec. 6.

## 2. RELATED WORK

Purcell et al. (2012) have conducted a large-scale study with 2,253 participants to investigate general web search. Among their most significant results are the facts that 65% are against using information from previous searches for search result ranking, 73% are against personalized search results and 68% are against targeted advertising. Furthermore, 91% stated they always or most of the time find their desired information, search engines are perceived as an unbiased source of information by 66% and more than half of the participants said that search results had gotten better over time. Also, positive search experience is more common than negative, although 38% stated that they have had the experience of feeling overwhelmed.

Each year, the *Consortium of Public Broadcasters in Germany* carries out a representative survey on Internet usage. The main findings of the 2014 edition (ARD/ZDF, 2014) are that in Germany, the fraction of people who go online (at least occasionally) has grown to 79.1% among those aged 14 and older and that it is mainly used by male users between 14 and 49 years. Moreover, usage duration per day has been increasing to 166 minutes in 2014 with mobile surfing and video consumption gaining importance. They also report that nowadays novel devices such as video game consoles or TVs are becoming considerable access points to the Internet. Some more detailed findings of this survey will be given along our own results.

While there has already been a considerable amount of research on mobile search (e.g., Carrascal and Church, 2015; Church and Oliver, 2011; Cui and Roto, 2008; Kamvar et al., 2009), the field of web search on novel devices is still in the early stages of development. For example, using a current search engine with novel input devices such as a video game controller still requires the use of an on-screen keyboard, which is highly cumbersome. Rather, interfaces should be provided that are tailored to the available input modalities, e.g., motion-sensing cameras or methods for voice recognition. Particularly *Apple* and *Google* have taken significant steps into the direction of leveraging speech as a new input modality for web search[5]. Other research has focused on the usage of Microsoft's *Kinect* (Liebling and Morris, 2012; Nebeling et al., 2014), 3D interfaces (Wiza et al., 2004), graspable interfaces (Keck et al., 2014) and image input (Both et al., 2014). Moreover, Wilson et al. (2010) evaluated existing search systems and provide a taxonomy intended for developers of future search interfaces.

Hearst (2011) points out four important trends that will drive future search interfaces. First, voice input will gain importance, as speaking is a more natural input technique than typing. Second, queries will become more natural language–like. Third, consumption of video content will increase, as it is less mentally demanding compared to reading. Fourth, social and collaborative search will gain importance and will be facilitated even across devices by frameworks. In general, Hearst predicts that search interfaces will become more natural.

All of these are promising research approaches with respect to the development of future search interfaces. Yet, a thorough investigation of the *user's perspective* is still missing. Also, there exist no *general requirements* that cover both, current *and* future search interfaces. For instance, Purcell et al. (2012) perform an empirical analysis of general web search but do not focus on different kinds of devices or interfaces while Wilson et al. (2010)—in their taxonomy for developers of future search interfaces—do not focus on an empirical analysis of users' perceptions and usage of current search systems.

[5] http://www.apple.com/ios/siri/, https://www.google.com/search/about/features/01 (July 20, 2016).

# 3. SURVEY

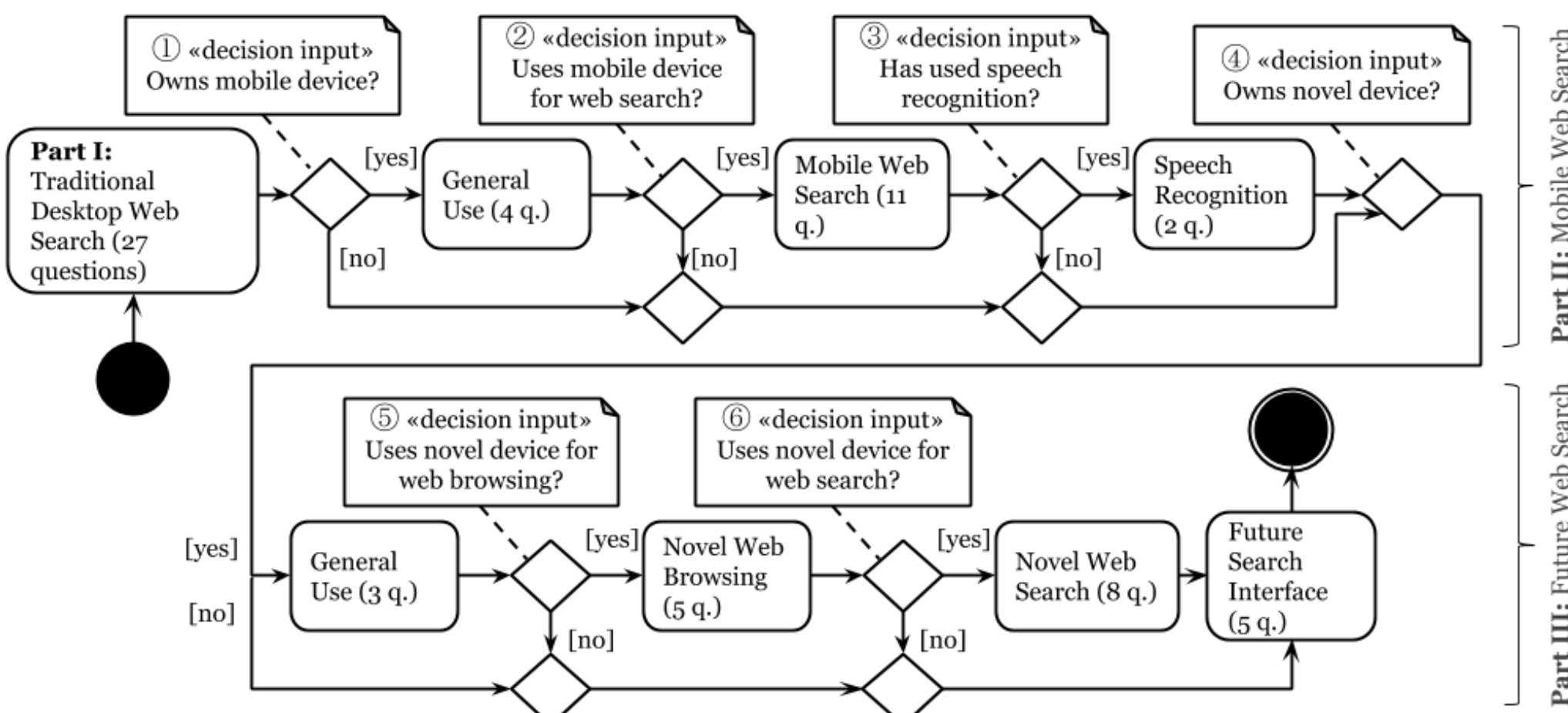

Figure 1. Flow chart of the survey with its different decision nodes. Questions for "no" answerers and demographic questions are omitted for simplicity

Our survey consisted of three main parts—*traditional desktop, mobile* and *future* web search—and a fourth part containing standard demographic questions (Fig. 1). This was necessary as we intend to provide a *general* set of requirements that spans the whole range from validating and revising existing search interfaces to supporting the development of new, "futuristic" ones. For answers expressing a frequency, likeliness or agreement with a given statement, we used 5-point Likert scales. We also posed a number of yes/no, multiple choice and open-ended questions. In total, the survey featured 98 questions of which 20 were optional.

The survey was prepared as a *Google Form* in German and English. Before publishing, it was revised by five experts from the search engine industry (one back-end developer, one front-end developer, one designer, one PhD student and one product manager). Participants were recruited through internal and external mailing lists, social networks and personal contact. Each participant who finished the survey received a thank-you gift in terms of a coupon worth 50 Euros for booking a trip on a German travel website. The complete survey and translated results are available via our online appendix at https://github.com/maxspeicher/ search-interface-requirements.

# 4. RESULTS

Overall, we collected 118 valid data sets. The participants (57.6% male, 42.4% female) had an average age of 29.8 years (σ=8.1) and were mainly German (N=103, 87.3%). Only seven participants (5.9%) stated they were not living in Germany. Furthermore, 114 participants (96.6%) stated they live in a German-speaking country. Hence, w.l.o.g. we assume that our results *refer to search interfaces for German-speaking users*. 23.7% of the participants were students, 55.1% had a formal qualification in computer science and 16.1% had one in web design (whereas only 1.7% stated they have a formal qualification in web design but none in computer science). 28.9% of the participants stated their work or studies are concerned with search engines. Due to the demographic structure of our test participants, ARD/ZDF (2014) will be the main source of comparison for our results (we consider the weighted average of the age groups 14–29 and 30–49), although it has to be considered that our audience was *more proficient than average*, which is reflected by the results. In general, our participants surfed the web for *more than 240 minutes (4 hours)* per day [ARD/ZDF (2014): 197 minutes], mainly for work (97.5%), news [75.4%, ARD/ZDF (2014): 55.2%], social networks (61.9%), shopping [41.5%, ARD/ZDF (2014): 18.6%] and videos [34.7%, ARD/ZDF (2014): 45.6%]. In the following, due to limited space, we are not able to provide in-depth results, but describe the ten key findings derived from participants' answers to our survey. From these findings we then derive the requirements specification that is the main contribution of this paper in Sec. 5. All results and qualitative feedback can be found in our online appendix.

Table 1. Main demographic properties of other studies we refer to

| | Purcell et al. (2012) | ARD/ZDF (2014) | Church and Oliver (2011) | Our study |
|---|---|---|---|---|
| N | 1,729* | 1,814 | 18 | 118 |
| Countries considered | U.S. | Germany | Multiple (Europe) | Multiple** |
| Languages | English, Spanish | German | English | German, English |
| Average age | n/a*** | n/a*** | 29 (σ=7.7) | 29.8 (σ=8.1) |

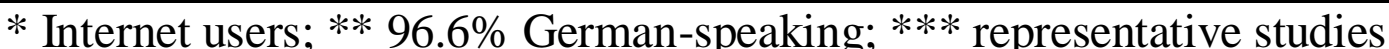
* Internet users; ** 96.6% German-speaking; *** representative studies

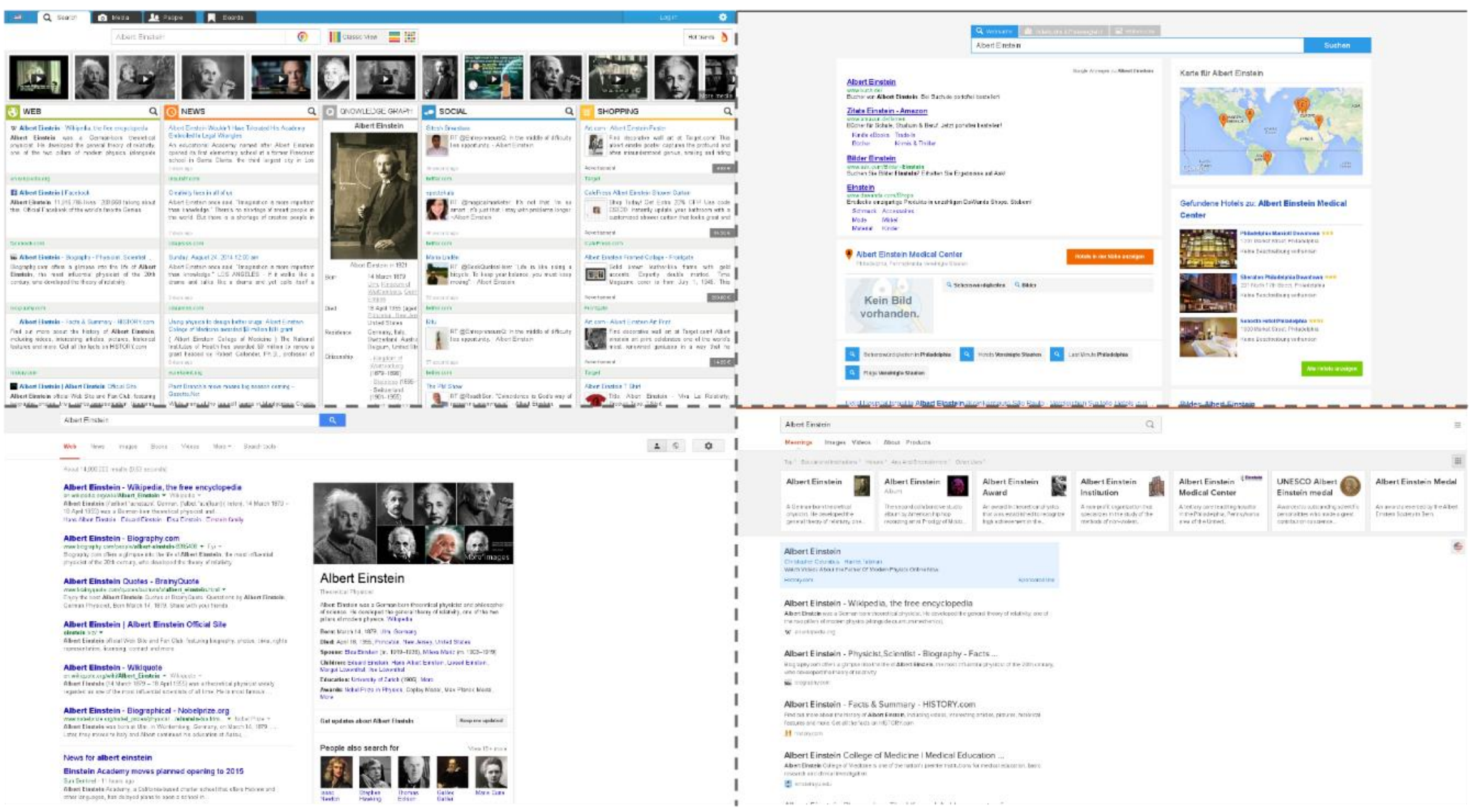

Figure 2. Clockwise from top left: *Qwant*, an experimental, non-public interface, *DuckDuckGo,* and *Google*

**Finding 1.** NUMBER OF CONSIDERED RESULTS PAGES. From two multiple-choice questions (I.10., I.11.)[6] we learned that in contrast to earlier research (e.g., Silverstein et al., 1998), users usually *do* consider pages beyond the first page of search results. If they still cannot find what they are looking for, they refine their search query rather than trying another search engine.

**Finding 2.** ADVERTISEMENTS. The results of six multiple-choice questions (I.4.–I.9.) show that users tolerate ads on SERPs if they are displayed in an unobtrusive way. Also, almost half of the users click on ads if they are helpful for their search. Our results indicate that acceptable numbers of ads are *two* above the search results and *five* elsewhere.

**Finding 3.** RESULTS PRESENTATION. From qualitative and quantitative assessments of different search engines (Fig. 2) by all 118 participants (I.13.–I.27), we found that users appreciate a clearly structured look & feel with a carefully balanced information density. They see advantages in the presentation of results from different domains, if they are relevant to their search query and not *just might* be of interest. Ditto for info boxes that answer queries directly on a SERP. Users tend to dislike scanning for information in two spatial dimensions at once (cf. Qwant.com).

**Finding 4.** MOBILE BROWSING. From 11 multiple-choice questions (II.5.–II.15.), we learned that mobile devices are widely accepted for web browsing and search. Also, users like to use familiar interfaces across devices, e.g., the Google *website* on mobile rather than a native app.

**Finding 5.** SPEECH RECOGNITION. Five multiple-choice questions (II.16.–II.20.) and additional qualitative feedback from 55 participants (46.6%) showed that in general, users are rather negative about speech recognition. Those who already used it are undecided about the importance of speech recognition in the future. Those who have not yet used it primarily feel uncomfortable talking to a device (particularly in public) and have strong privacy concerns.

[6] These numbers refer to questions from our survey, which can be looked up at http://goo.gl/u7YvCv.

**Finding 6.** NOVEL BROWSING. Based on six multiple-choice questions and qualitative feedback (III.5.–III.8., III.21.–III.24.) from 53 participants (44.9%), we found that using a novel device for web browsing is not very common among users, which seems to be mainly due to the different use cases. Also, users are not satisfied with the available input modalities. Currently, there seems to be no alternative that can cope with the efficiency of regular keyboard input.

**Finding 7.** USE CASES. Qualitative feedback from 77 participants (65.3%) indicated that particularly, the majority of users see no reason for mixing different use cases. This is in line with ARD/ZDF (2014), who found that all-in-one devices are far from being the norm. About 30% of our participants owned at least three web-enabled devices. ARD/ZDF (2014) report an average of 5.4 web-enabled devices per household, of which, however, only 2.8 are used.

**Finding 8.** CREATURES OF HABIT. A total of 27 multiple-choice questions (e.g., II.7., II.22., III.18., III.21.) as well as qualitative feedback from 113 participants (95.8%) indicated that users prefer to use systems they already know. This accounts for interfaces (e.g., Google) as well as devices (e.g., keyboards, touchscreens).

**Finding 9.** POTENTIAL OF NOVEL SEARCH. From 22 multiple-choice questions and qualitative feedback (III.2.–III.28.) from 77 participants (65.3%), we learned that users generally acknowledge the potential of mobile and novel devices and input modalities for web search (and vice versa). Although our participants were generally satisfied with mobile search experience, they noticed plenty of room for improvement concerning search interfaces beyond desktop PCs. The latter particularly applies to interfaces for novel devices.

**Finding 10.** Based on qualitative feedback from 30 participants (25.4%), A FUTURE SEARCH INTERFACE realized in the midterm would involve *touch interaction*, *speech recognition* and *familiar technology* that is deployed *across devices*. For example, the Google search, running on a Smart TV that accepts gesture and keyboard input from an iPad. At home, the distributed search interface would be augmented with optional voice input. In public, the interface would be available via a single mobile device.

# 5. REFOCUS: REQUIREMENTS SPECIFICATION

In the following, we present a first draft of *REFOCUS*—our novel set of general requirements, which is directly inferred from the findings of the exploratory study above. Yet, to guarantee the generality of the proposed requirements (w.r.t. German-speaking users) as well as their validity for current and future search interfaces, we had them reviewed by 12 experts working in the search engine industry. This happened following the *find–fix–verify* pattern by Bernstein et al. (2010), however, with a combined *fix–verify* phase.

## 5.1 Intermediate Results (*Find* Phase)

Current and future search interfaces for German-speaking users should (information in square brackets refers to the validation from Sec. 5.2):

1. *Finding 1* → provide adequate ranking also for results beyond the initial viewports. If the user cannot find what they are looking for, proper related search inputs should be proposed. [*U*niversally valid]
2. *Finding 2* → provide ads in a subtle way. They should accompany the search results rather than demanding their space in the initial viewport. If ads are placed above results, their number should not exceed *two*; if they are placed elsewhere, their number should not exceed *five*. [*C*urrent search interfaces]
3. *Finding 3* → above all ensure that all displayed information is relevant to the search query. The most relevant piece of information should be immediately identifiable, as form follows function. If applicable, semantic results should answer the query directly in the initial viewport. [*U*]
4. *Findings 3, 8* → not require the user to scan information in more than one spatial dimension simultaneously. Different categories of results should be selectable rather than displaying all information at once. [*U*]

5. *Findings 3, 9* → make proper use of the available space, particularly with respect to increasing display sizes. [*U*]
6. *Findings 4, 6, 8* → leverage the advantages of interfaces familiar to the user. Transition to radically new concepts should happen in small steps. Fallback functionality for legacy input/output modalities should be provided. [*U*]
7. *Findings 5, 6, 8* → always provide an easily reachable alternative to voice input. Users should be properly informed about privacy issues and where it is unproblematic to use speech recognition. [*U*]
8. *Findings 6, 7* → be tailored to the target audience of the devices they run on. For instance, a web-enabled video game console should provide a search interface that focuses on finding games and information relevant to those. Furthermore, it must be optimized for efficient input with the same device as is used for playing. [*U*]
9. *Findings 9, 10* → support cross-device interaction, i.e., the search interface is distributed and input/output can optionally happen on different devices. [*F*uture search interfaces]
10. *Findings 9, 10* → be ubiquitous, thus focusing on devices users carry around most of the time. These include, but are not restricted to, mobile devices, devices like Google Glass and fitness bracelets. [*F*]
11. *Finding 10* → consider input/output formats other than text, such as images, music, gestures, 3D visualizations and image-based zoomable user interfaces. [*F*]

## 5.2 Reviewed & Validated Requirements (*Fix–Verify* Phase)

The above intermediate results were given to 12 dedicated experts in fields related to the development of search engines and corresponding interfaces. Nine of them worked in academia, two worked in industry and one in both. Four considered themselves to be a practitioner, one to be a researcher and seven said "half/half". 11 were male (one female) at an average age of 29.7 (σ=3.70), with 11 coming from Germany and one from Spain (but living in Germany). Seven experts owned a Master's, one a Bachelor's and one a PhD degree; three said "other" (e.g., the German *Diplom,* which is equal to a Master's degree).

Table 2. Experts' ratings of the set of requirements concerning their validity for different types of search interfaces

| Requirement | Universally valid | Current search only | Future search only |
|---|---|---|---|
| R1 | 11 | 0 | 1 |
| R2 | 6 | 6 | 0 |
| R3 | 10 | 1 | 1 |
| R4 | 9 | 2 | 1 |
| R5 | 10 | 0 | 2 |
| R6 | 10 | 2 | 0 |
| R7 | 7 | 1 | 4 |
| R8 | 7 | 3 | 2 |
| R9 | 6 | 0 | 6 |
| R10 | 3 | 1 | 8 |
| R11 | 5 | 0 | 7 |

They were first instructed to thoroughly read the detailed results of the empirical study as well as the ten key findings. After that, each expert was asked to review the set of requirements and for each one indicate whether it is (1) universally valid, (2) valid only for current search interfaces or (3) valid only for future search interfaces. Also, they had to indicate whether the requirement should be removed, should stay in its current form or needs adjustments. If they indicated the requirement should be altered, they had to state how.

The experts found seven requirements to be universally valid (*U*), one to be valid only for current search interfaces (*C*) and three to be valid only for future search interfaces (*F*). This was determined by majority vote (>6) and is summarized in Tab. 2 and moreover indicated in the list above in square brackets. In the cases of R2 and R9, six votes sufficed for "current search only" and "future search only", respectively, because votes for "universally valid" implicitly count towards those as well.

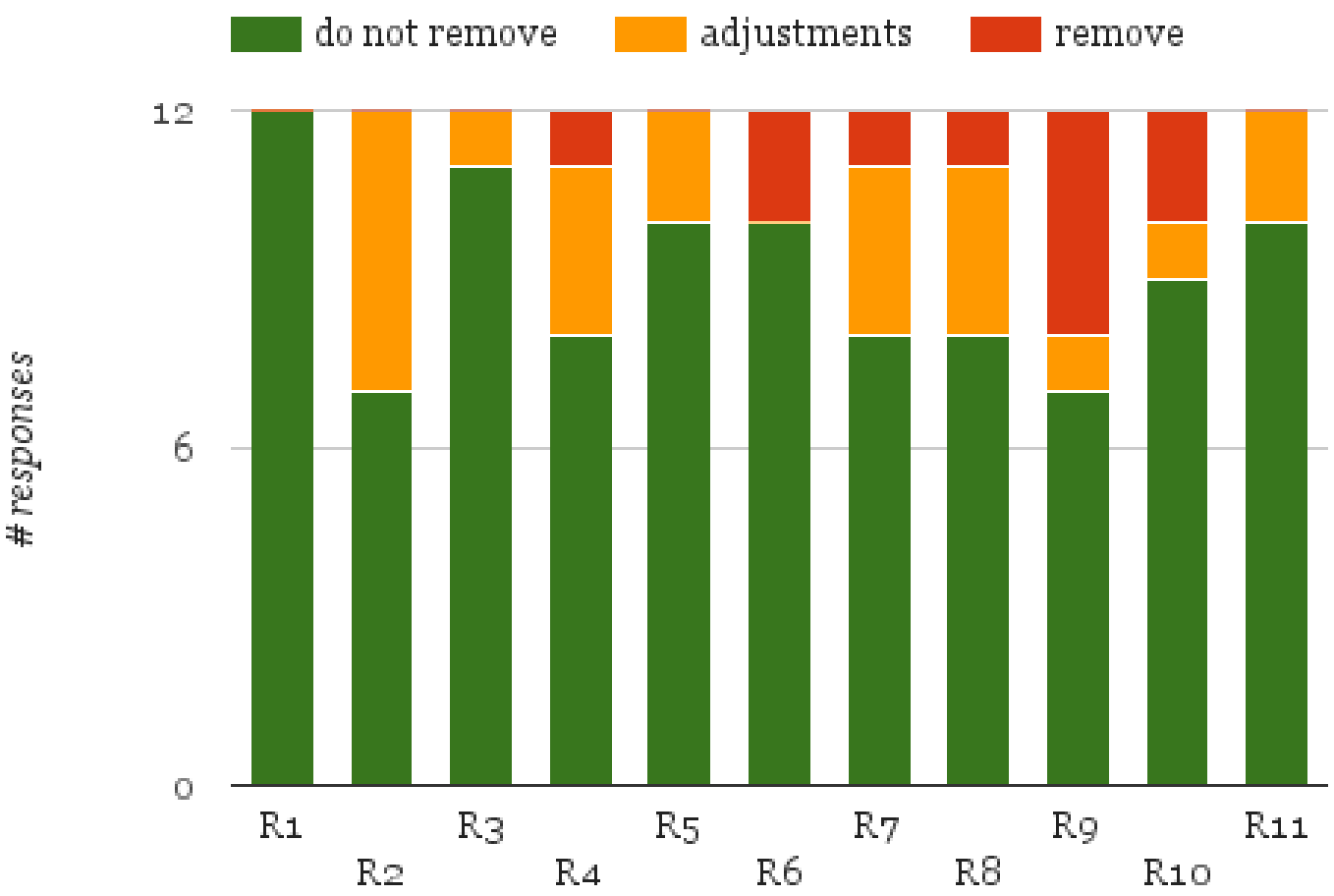

Figure 3. Experts' votes concerning whether requirements should be adjusted or not

As for the adjustment of requirements, we defined the following rule beforehand: If the majority of experts (>6) vote for revising a requirement and at least two experts propose the same adjustment, the requirement will be changed accordingly. Yet, the number of votes for "adjustments" did not exceed the threshold of six for any of the requirements. Besides, regarding the adjustments that were proposed, none was requested by more than one expert. Hence, the set of requirements named REFOCUS can be considered valid in their current form w.r.t. the respective type(s) of interface(s)—universal, current or future.

When looking at the final set of requirements, one could argue that they are formulated in a rather vague way and contain recommendations that are well-known and have been long established. At this point, we want to note that the intention of our *exploratory* study was to provide a set of requirements that are of a *general* nature and *agnostic* w.r.t. devices and input/output modalities. This implies that our requirements have to be formulated in a non-specific way. Hence, they should be seen as a mixture of existing (which are re-validated by our study) and novel findings (which pave the path towards future search interfaces).

The complete survey and detailed results for the above follow-up expert study can be found in our online appendix.

## 6. DISCUSSION & CONCLUSIONS

We have reported on the results of an exploratory user study involving 118 participants. The underlying survey was—to the best of our knowledge—the first to be concerned with use and perception of and expectations towards current *and* future search interfaces. Our results show that users are generally satisfied with the current state of desktop and mobile search, but are skeptical towards novel devices beyond mobile when it comes to web search. Also, keyboard and touch input are preferred over new input modalities such as speech recognition and motion-sensing technology. This is mainly due to users sticking to interfaces and technologies they are familiar with. From our ten key findings, we were able to deduce *REFOCUS*—a novel 11-point *requirements specification for current and future search interfaces for German-speaking users*, which is particularly intended for search engine developers and designers and in an additional step has been reviewed and validated by 12 experts.

The results of our exploratory study give a good impression of how users perceive current search interfaces as well as established and emerging technologies for web browsing. During analysis, it turned out that results became more and more *qualitative* as the survey progressed from the "Traditional Desktop Web Search" to the "Future Web Search" part. This was the case due to the investigated technologies becoming more and more advanced, which meant that the corresponding questions were applicable only to a thinning fraction of our participants. On the one hand, a main reason for this was the relatively small size of our overall sample (N=118). Yet, our audience shows demographic structures that are partly similar to established surveys (e.g., ARD/ZDF, 2014). On the other hand, it reflects the fact that novel technologies for web search have not yet arrived to take their place at the center of society. Also, this gave us the opportunity to more thoroughly review individual opinions regarding novel technologies and approaches to web search.

This has proven valuable for understanding users' expectations towards future search interfaces. The qualitative statements given can be considered profound as our audience was *more proficient* in web design and search technology than the average user. However, as part of our future work, we intend to carry out the survey with a larger audience at a later point in time to investigate trends in web browsing/search on novel devices. Moreover, we want to extend the part on future web search and consider it in isolation with an even more specifically chosen group of participants for a more quantitative analysis.

To conclude this paper, we want to give a *vision of the future of search interfaces*. In an open-ended question, we asked the participants of our study to share their visions of a potential search interface of the future. Combining their responses, a future search interface realized in the long term would be *ubiquitous*, involving *devices such as Google Glass* as well as *eye tracking*. Interfaces would be directly projected into the user's field of view. *Image input* (e.g., Both et al., 2014) would be possible by simply focusing things. Optimally, such interfaces would be further augmented by *brain–computer* interaction technology (cf. del R. Millan, 2006). Imagine yourself lying on your couch wearing your augmented reality glasses. On TV you see an interesting building, so you look at the "Pause" symbol in the top-left corner of your field of view. Your TV stops and by focusing it, your glasses automatically perform a search for more information about the building. The results are visualized in 3D right in your living room. They are sorted by date on the X axis, by popularity on the Y axis and by type (image, news, blog etc.) on the Z axis. By focusing different parts of the visualization you zoom into and explore the results and finally learn that the interesting building is the famous Sagrada Família in Barcelona, which has to be completed yet.

# ACKNOWLEDGEMENT

Many thanks go to all survey participants! We also thank F. Funke, C. Lemke, S. Nitsch, M. Röder and S. Schattenberg for their help with creating the survey. This work has been supported by the ESF and the Free State of Saxony.

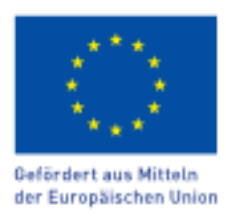

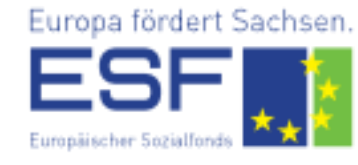